\documentclass[conference]{IEEEtran}

\usepackage[utf8]{inputenc}
\usepackage[T1]{fontenc}

\IEEEoverridecommandlockouts
\usepackage{cite}
\usepackage{amsmath,amssymb,amsfonts}
\usepackage{algorithmic}
\usepackage{graphicx}
\usepackage{textcomp}
\usepackage{xcolor}
\usepackage{amsthm}
\newtheorem{definition}{Definition}
\usepackage{multirow}
\usepackage{booktabs, caption}
\usepackage{xspace}
\newcommand{\sys}{{EidosDoc}\xspace}

\def\BibTeX{{\rm B\kern-.05em{\sc i\kern-.025em b}\kern-.08em
    T\kern-.1667em\lower.7ex\hbox{E}\kern-.125emX}}
\begin{document}

\title{EidosDoc: Implicit Structure Encoding for Cost-Effective Semi-Structured Document QA\\
}

\author{\IEEEauthorblockN{Teng Lin}
\IEEEauthorblockA{\textit{DSA Thrust} \\
\textit{HKUST(GZ)}\\
Guangzhou, China \\
tlin280@connect.hkust-gz.edu.cn}
\and
\IEEEauthorblockN{Yuyu Luo}
\IEEEauthorblockA{\textit{DSA Thrust} \\
\textit{HKUST(GZ)}\\
Guangzhou, China \\
yuyuluo@hkust-gz.edu.cn}
\and
\IEEEauthorblockN{Nan Tang}
\IEEEauthorblockA{\textit{DSA Thrust} \\
\textit{HKUST(GZ)}\\
Guangzhou, China \\
nantang@hkust-gz.edu.cn}

}

\maketitle

\begin{abstract}
Semi-structured documents are ubiquitous in scientific reports, financial statements, and technical manuals. Question answering over such documents requires simultaneous understanding of text, tables, charts, and complex hierarchical layouts. Existing methods either rely on repeatedly calling large language models for structure parsing and retrieval, leading to high cost and large latency, or they flatten the document and lose layout and hierarchy information, sacrificing answer accuracy. 
To address this, we propose \sys, a novel system that achieves state-of-the-art accuracy with minimal computational expense. Our approach introduces three core innovations. (1) An Implicit Structure Encoder trained via contrastive learning and a structure consistency loss. This module jointly embeds hierarchical relationships, spatial positions, and textual content into a dense vector space, capturing document structure holistically without the need for manually defined and error-prone constructions. (2) A Hybrid Retrieval Pipeline that leverages BM25, layout fingerprints, and a lightweight cross-encoder to perform high-precision retrieval entirely without invoking an LLM, drastically reducing cost and latency. (3) A Dynamic Evidence Expansion mechanism that adaptively retrieves spatially adjacent and structurally related evidence, overcoming the evidence omission common in fixed-path retrieval methods.
We evaluate \sys on four benchmarks, and comprehensive evaluations show that \sys achieves a new state-of-the-art accuracy on the four benchmarks. Crucially, it does so with a 50× reduction in cost and 4× lower latency compared to the previous state-of-the-art Method. These results demonstrate that \sys establishes a new optimal trade-off among accuracy, cost, and speed, offering a practical and scalable path for accurate semi-structured document analysis.
\end{abstract}


\section{Introduction}

\begin{figure*}[htbp]
\centering
\includegraphics[width=0.91\linewidth]{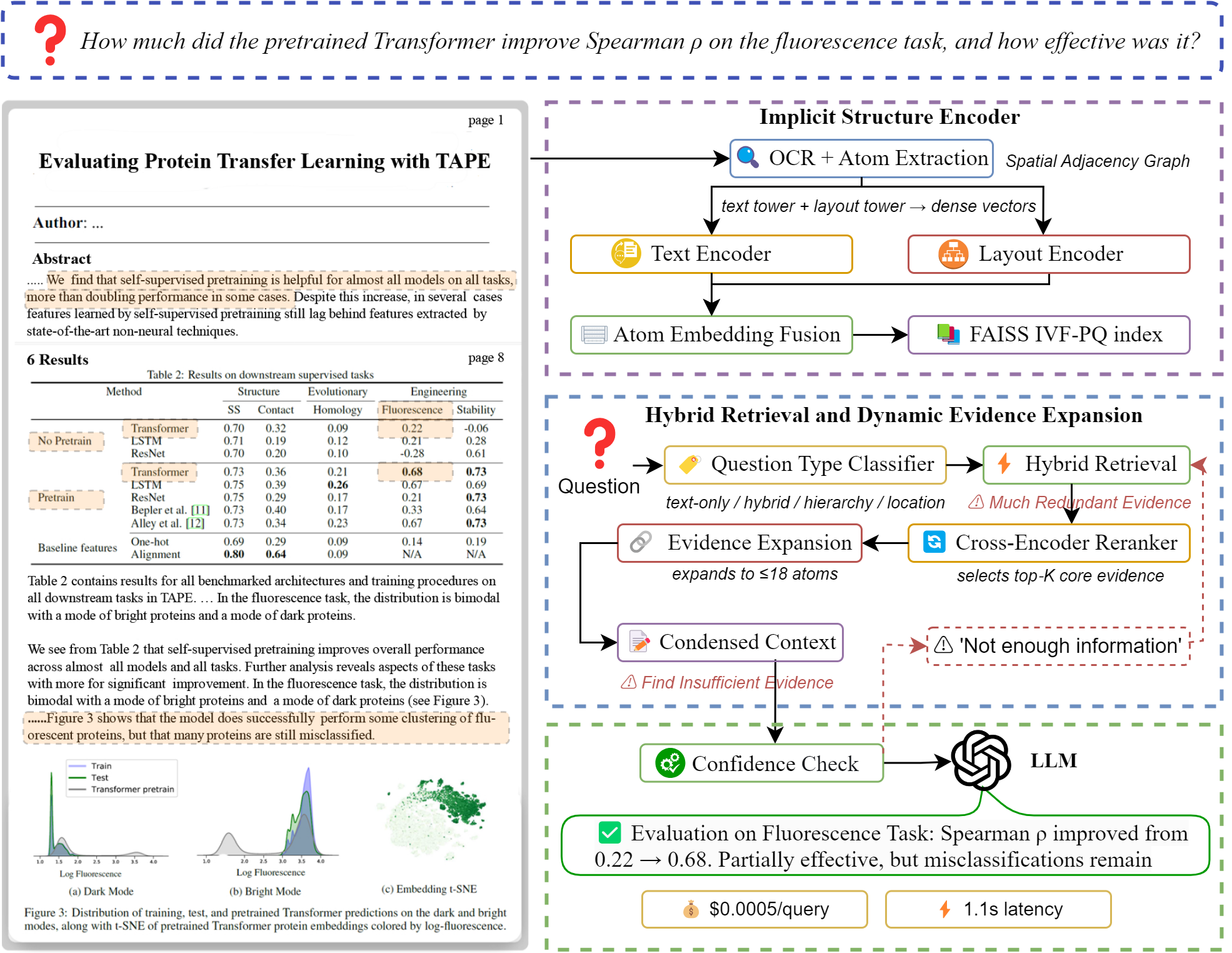}
\caption{Overview of the \sys pipeline and its output on an example question, illustrating the key stages and core modules.}
\label{fig:pipeline}
\end{figure*}

Semi-structured documents, such as scientific reports, financial statements, technical manuals, and government forms, are pervasive across real-world applications. Unlike plain text or rigidly structured databases, these documents integrate diverse elements, including paragraphs, tables, charts, images, multilevel headings, sidebars, and footnotes, often arranged in complex and irregular layouts~\cite{10.1145/3768156, EVAPORATE, tang2023unifying, lin2025Simplifying}. Answering natural language questions over such documents is both practically valuable and technically challenging~\cite{MoDora}. For example, as illustrated in Figure~\ref{fig:pipeline},  consider a machine learning paper~\cite{NEURIPS2019_37f65c06} containing Table~2 reporting Spearman correlation coefficients for LSTM and Transformer on a fluorescence prediction task, Figure~3 showing the predicted vs.\ true values for both models. A question \textit{``How much did the pretrained Transformer improve Spearman $\rho$ on the fluorescence task, and how effective was it?''} requires the system to: (i) extract the numerical improvement from the table (e.g., from 0.22 to 0.68); (ii) interpret the figures to understand qualitative effectiveness (e.g., tighter clustering, lower variance); (iii) synthesize the quantitative and qualitative evidence; and (iv) produce a comparative answer. Such multi-step, cross-component reasoning is typical for semi-structured document analysis~\cite{lin2026montecarlotreesearch}.

A large body of research has addressed document understanding, but existing methods exhibit critical shortcomings when applied to semi-structured document QA. Content extraction methods~\cite{QUEST, EVAPORATE, liu2025palimpzest, saad-falcon-etal-2024-pdftriage, sadia-etal-2025-squid,lin2026annoretrieveefficientstructuredretrieval} transform documents into relational tables or attribute-value pairs. While efficient for pure text, they discard layout and structural information, making them unable to answer questions that depend on spatial or hierarchical cues. Structure extraction methods~\cite{zendb, lin2026structurethenquery, sun-etal-2025-docagent,lin2026docsageinformationstructuringagent} attempt to recover hierarchical organizations such as chapters and page mappings. However, they typically operate at the page level or rely on brittle visual clustering, leading to incorrect or flattened hierarchies. Moreover, they often fail to model non-textual elements (tables, charts) and their relations with surrounding text. End-to-end multimodal models~\cite{LayoutLMv3, DocOwl2, achiam2023gpt, DocFormer} process documents as images and answer questions in a single pass. They suffer from high computational cost, limited context length, and poor fine-grained text understanding. In particular, they frequently overlook cross-element relationships, such as aligning a chart with its referring paragraph. Retrieval-augmented generation (RAG) methods~\cite{SVRAG, M3DocRAG, lin2025srag, textRAG, yang2025superrag, cao-etal-2025-neusym, lin2025structured} retrieve relevant chunks or page images using embeddings. They lose layout structure and often fail to perform multi-step reasoning that requires bridging information across different document regions. Tree-based structured approaches~\cite{MoDora} represent a recent advance. They construct explicit component-correlation trees and use LLMs for node selection, summarization, and verification. While achieving state-of-the-art accuracy, they incur high cost and high latency  due to multiple LLM calls during retrieval. Furthermore, the tree construction itself is fragile, and errors in hierarchy detection or OCR propagate and cause wrong or missed evidence. In summary, existing methods face a fundamental trade-off: high accuracy comes at the expense of high cost and latency, while low-cost solutions sacrifice accuracy by flattening or ignoring structure~\cite{DBLP:journals/corr/abs-2405-04674}.

The core challenges of existing methods for semi-structured document question answering can be summarized as follows. First, elements extracted by OCR are often fragmented and detached from their semantic context. A table cell or a chart title may be isolated from the surrounding explanatory text, making it difficult to interpret them in isolation~\cite{MoDora}. Second, existing methods lack effective representations to capture hierarchical structures (e.g., associating a table with nested chapter titles) and to preserve layout-specific distinctions (e.g., differentiating a sidebar from the main content). Third, answering complex questions frequently requires retrieving and aligning evidence scattered across multiple regions or pages, such as linking a descriptive paragraph to a chart located elsewhere in the document~\cite{gong2025mhierrag,LADRAG}.

We propose \textbf{\sys}, a novel system for semi-structured document analysis that simultaneously achieves higher accuracy, lower cost, and lower latency than the current state of the art. \sys abandons explicit tree construction and LLM-driven retrieval. Instead, it introduces three core innovations. First, an \textbf{implicit structure encoder}. We train a dual-encoder architecture (lightweight Transformer) using contrastive learning and a novel structure consistency loss. The encoder jointly embeds text, layout (bounding boxes, page indices), and hierarchical relationships (e.g., parent-title associations) into a dense vector space. This completely avoids building an explicit document tree, which is brittle and expensive. Second, a \textbf{hybrid retrieval pipeline}. During query processing, retrieval is performed using a combination of BM25, layout fingerprinting, dense vector similarity, and a lightweight cross-encoder reranker. No large language model is invoked during retrieval; all components run locally with negligible cost. Third, a \textbf{dynamic evidence expansion mechanism}. Instead of following a fixed tree path (as in prior work), \sys dynamically expands evidence atoms based on spatial adjacency and structural relatedness. This allows the system to recover cross-region evidence that would be pruned by tree-based methods, significantly improving recall. Finally, answer generation is performed by a single call to a LLM on a highly condensed context.

This paper makes the following contributions.
\begin{itemize}
    \item \textbf{Implicit structure encoding}: We propose a novel training paradigm that embeds hierarchical and layout information into dense vectors without explicit tree construction, using contrastive learning and structure consistency loss.
    \item \textbf{Hybrid retrieval pipeline}: We design a lightweight, fully local retrieval pipeline that achieves high recall and precision using only BM25, layout fingerprints, dense vectors, and a small cross-encoder, reducing cost and latency compared to LLM-based retrieval methods.
    \item \textbf{Dynamic evidence expansion}: We introduce a mechanism that dynamically retrieves spatially and structurally related evidence atoms, overcoming the evidence omission problem of fixed-path tree retrieval.
    \item \textbf{State-of-the-art empirical results}: On four benchmarks, \sys consistently outperforms prior SOTA and leading across all question types. It surpasses the previous best methods while reducing per-query cost by over 50× and latency by nearly 4×. Extensive ablations validate the contribution of each component.
\end{itemize}

\begin{figure*}[htbp]
\centering
\includegraphics[width=1\linewidth]{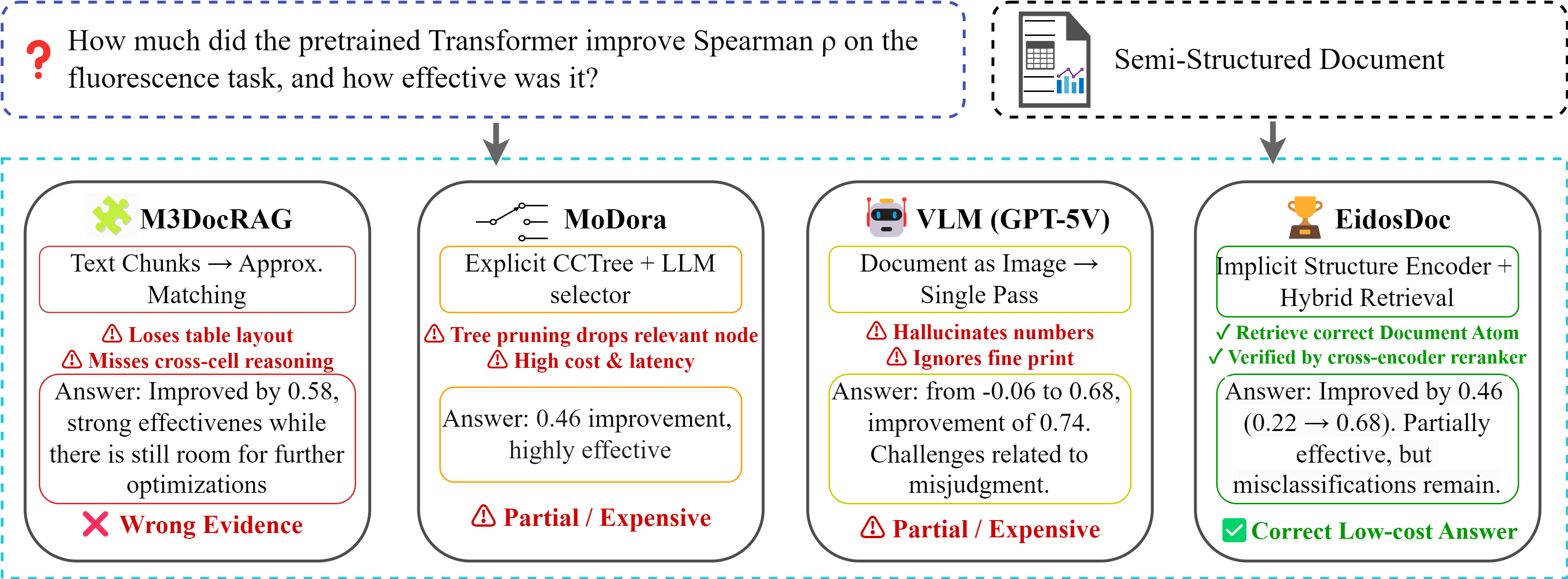}
\caption{A comparison of different document processing methods answering a specific question about protein fluorescence. While M3DocRAG provides wrong evidence and MoDora/VLMs are costly or prone to errors, only \sys get the correct numerical improvement (0.46) and faithfully describes the effectiveness.}
\label{fig:related}
\end{figure*}
\section{Preliminaries}

In this section, we establish the basic concepts, notation, and problem definitions used throughout this paper. We also briefly characterize the limitations of prior work that motivate the design of \sys. 

\subsection{Document Representation}

A semi-structured document is a visual artifact that conveys information through a mixture of textual and non-textual elements arranged in a complex layout. Examples include scientific reports, financial statements, technical manuals, and government forms. Unlike plain text, which is a linear sequence of characters, or fully structured data, which adheres to a rigid schema, a semi-structured document contains paragraphs, titles, tables, charts, images, sidebars, headers, footers, and page numbers. These elements are not independent; they interact through spatial proximity, hierarchical grouping (e.g., sections and subsections), and cross-references (e.g., ``as shown in Table~3'').

To enable fine-grained retrieval and reasoning, we introduce the concept of a \textit{document atom} as the fundamental unit of information.

\begin{definition}[Document Atom]
A document atom is a tuple \(a = (\textit{type}, \textit{content}, \textit{bbox}, \textit{page}, \textit{meta})\), where:
\begin{itemize}
\item \(\textit{type} \in \{\textit{text}, \textit{title}, \textit{table\_cell}, \textit{chart\_caption}, \textit{image},  \textit{supplement}\}\) specifies the kind of content.
\item \(\textit{text}\) is a string. For textual atoms (text, title, table cell, chart caption, supplement), it is the extracted text. For non-textual atoms (image), it is a short description generated by a lightweight model.
\item \(\textit{bbox} = (x_1, y_1, x_2, y_2)\) is the bounding box normalized to page coordinates, where \((x_1, y_1)\) is the top-left corner and \((x_2, y_2)\) the bottom-right corner.
\item \(\textit{page}\) is the page number (starting from 1).
\item \(\textit{meta}\) is a set of structural hints, including:
  \begin{itemize}
  \item \(\textit{depth} \in \{0,1,2,3,4,5\}\) (estimated heading level; 0 for non-title atoms);
  \item \(\textit{is\_title} \in \{0,1\}\);
  \item \(\textit{parent\_embedding}\): a 128-dimensional vector approximating the embedding of the atom’s parent title (or zero vector if none).
  \end{itemize}
\end{itemize}
These meta attributes are predicted by a small auxiliary model that is trained on a small corpus of annotated documents. The model does not construct an explicit tree; it only produces per-atom estimates.
\end{definition}

The key property of an atom is that it is both \textit{self-contained} (it can be interpreted independently) and \textit{granular} (it does not aggregate heterogeneous content). For example, a table is not stored as a single atom; each cell is an atom. The relationship between cells in the same row or column is captured not by explicit grouping but by spatial adjacency (see below). This design avoids the brittleness of explicit structure extraction.

\begin{definition}[Document Representation]
A document \(D\) is represented as a set of atoms \(\mathcal{A} = \{a_1, a_2, \dots, a_n\}\) together with a \textit{spatial adjacency graph} \(G_s = (V, E)\), where \(V = \mathcal{A}\) and an undirected edge \((a_i, a_j) \in E\) exists if and only if:
\begin{itemize}
\item \(a_i.\textit{page} = a_j.\textit{page}\); and
\item the bounding boxes of \(a_i\) and \(a_j\) either overlap or are within a small vertical/horizontal distance \(\delta\) (we use \(\delta = 0.05\) in normalized coordinates).
\end{itemize}
\end{definition}

The spatial graph captures local layout relationships (e.g., two cells in the same row, a caption below a chart) without imposing a hierarchical tree. It is used only during retrieval to expand evidence, as explained in later sections.

\subsection{Problem Formulation}

We study the task of \textit{semi-structured document question answering} (QA). Given a document \(D\) (represented as atoms \(\mathcal{A}\) and spatial graph \(G_s\)) and a natural language question \(Q\), the system must produce an answer \(A\) that is correct according to the information in \(D\).

\begin{definition}[Semi-Structured Document QA]
Let \(D\) be a semi-structured document with atom set \(\mathcal{A}\) and spatial graph \(G_s\). Let \(Q\) be a question expressed in natural language. The task is to compute a function \(f: (D, Q) \mapsto A\), where \(A\) is a string (which could be a short phrase, a number, or a date) such that:
\begin{itemize}
\item There exists a set of relevant atoms \(\mathcal{R} \subseteq \mathcal{A}\) that collectively contain the information needed to answer \(Q\);
\item \(A\) is derived from \(\mathcal{R}\) (possibly through reasoning that combines multiple atoms);
\item No information outside \(\mathcal{R}\) is necessary to correctly answer \(Q\) (i.e., \(\mathcal{R}\) is a minimal sufficient set).
\end{itemize}
\end{definition}

This definition emphasizes that the system must first \textit{retrieve} the right evidence (\(\mathcal{R}\)) and then \textit{reason} over it to produce the answer. The quality of the answer thus depends critically on the recall and precision of the retrieval step.

\paragraph{Question Typology.}
To facilitate systematic evaluation, prior benchmarks~\cite{M3DocRAG, dude, DocVQA, MoDora} and real-world use cases identify four broad categories of questions. A robust system must handle all of them.

\begin{itemize}
\item \textbf{Text-only questions} can be answered using only plain text atoms (i.e., type = text or title). Example: ``What is the sample size in Experiment 2?'' The answer can be extracted from a paragraph without consulting tables or charts.

\item \textbf{Hybrid questions} require information from non-textual atoms, such as table cells or chart captions. Example: ``Which group had the lowest feather score in week 3?'' The system must locate the correct table or chart and extract the relevant cell value.

\item \textbf{Hierarchy-aware questions} depend on the document’s section structure. Example: ``What are the two sub-topics discussed under Section 4.1?'' The answer is a list of subsection titles, which requires understanding that Section 4.1 has children at a deeper depth.

\item \textbf{Location-aware questions} rely on spatial position. Example: ``What is the value in the bottom-left cell of the first table?'' The system must interpret ``bottom-left'' as a region within the table’s bounding box and identify the corresponding atom.
\end{itemize}

More complex multi-hop questions (e.g., ``Compare the feather score of Decoquinate in winter to that in summer'') can be decomposed into sequences of the above types using existing query decomposition techniques. Our work focuses on the core retrieval and reasoning challenges; we do not assume any particular decomposition method but note that our system can be combined with them.

\subsection{Limitations of Prior Work}

A large body of research has addressed document understanding. However, as shown in Figure~\ref{fig:related}, existing methods exhibit systematic limitations when applied to semi-structured document QA. 

\paragraph{Retrieval-Augmented Methods}
Recent RAG approaches to semi-structured document understanding vary in retrieval granularity and structural awareness. Traditional methods like TextRAG~\cite{textRAG} treat documents as flat text chunks, discarding layout and hierarchy, and cannot answer location-aware questions. Page-level methods such as M3DocRAG~\cite{M3DocRAG} and SV-RAG~\cite{SVRAG} retrieve entire page images using multi-modal models, preserving visual cues but suffering from coarse granularity and noise. More advanced approaches build explicit structures: SuperRAG~\cite{yang2025superrag} constructs a layout-aware knowledge graph, LAD-RAG~\cite{LADRAG} builds a symbolic document graph with cross-page dependencies. Despite these advances, all RAG methods share common limitations: retrieval quality bounds answer accuracy, explicit structure extraction is brittle, and invoking LLMs during retrieval incurs high cost and latency. These limitations motivate EidosDoc, which eliminates LLM calls from retrieval while preserving structure through implicit encoding and dynamic expansion.

\paragraph{Explicit structure methods}
To preserve document structure, several works build explicit hierarchical representations. ZenDB~\cite{zendb} uses visual feature clustering and LLMs to detect titles and their associated body text, forming a shallow tree. DocAgent~\cite{sun-etal-2025-docagent} extracts an XML-based outline to guide agent actions. MoDora~\cite{MoDora} constructs a Component-Correlation Tree (CCTree) through local alignment aggregation and bottom-up summarization, and then uses LLMs for node selection, verification, and answer generation. TWIX~\cite{lin2025twix} reconstructs structured data from templatized documents by inferring the underlying visual template. These methods achieve state-of-the-art accuracy on semi-structured QA. However, they have critical drawbacks. (i) High cost and latency. Because they invoke LLMs multiple times during retrieval (e.g., for node selection and verification), their per-query cost is high and latency is long. This makes them impractical for large-scale or interactive applications~\cite{lin2025lightkggsimpleefficientknowledge,lin-etal-2025-mebench}. (ii) Brittleness. Explicit tree construction depends on accurate OCR and reliable heuristic rules (or LLM-based classification) for grouping elements and detecting hierarchy. Errors at this stage propagate and can cause missing or misplaced nodes. Moreover, the fixed tree path may prune evidence that is relevant but not reachable via the pre-defined hierarchy.(iii) LLM dependency for retrieval. Using LLMs not only for generation but also for retrieval decisions introduces unpredictable behavior and high variance.

\paragraph{End-to-end multimodal LLMs}
Models such as GPT-5V~\cite{achiam2023gpt}, DocOwl2~\cite{DocOwl2}, and LayoutLMv3~\cite{LayoutLMv3} process document pages as images (sometimes with OCR tokens). They attempt to answer questions directly without explicit retrieval. While they avoid the need for separate retrieval components, they suffer from their own limitations: (i) they have limited context windows (typically a few thousand tokens, which corresponds to only a few pages); (ii) they are expensive because each page is encoded as a high-resolution image; (iii) they often miss fine-grained textual details because visual representations are inherently lossy; (iv) they are not robust to multi-page reasoning because they cannot easily reference information from non-adjacent pages. Consequently, their accuracy on semi-structured documents is far below that of tree-based methods, and their cost is even higher.

\section{System Overview}

As shown in Figure~\ref{fig:pipeline}, \sys is a three-stage pipeline for semi-structured document question answering that achieves high accuracy, low cost, and low latency by abandoning explicit tree construction and LLM-driven retrieval. The system first performs offline preprocessing once per document, then executes an online query processing pipeline that uses only lightweight local models, and finally generates an answer with a large language model(LLM). The following paragraphs provide an overview of each stage; detailed explanations of the core components (implicit structure encoder, hybrid retrieval, dynamic evidence expansion, and answer generator) are provided in later sections.

\textbf{Offline Preprocessing.} In this stage, the raw document is transformed into a searchable representation without any LLM involvement. A lightweight OCR engine extracts all text fragments, table cells, chart captions, and auxiliary elements; each becomes a document atom with its type, text, bounding box, and page number. A small language model predicts structural metadata for each atom, including a depth estimate and a title flag. Using bounding boxes and page numbers, the system builds a spatial adjacency graph where edges connect atoms that are on the same page and whose bounding boxes overlap or are within a small threshold. Finally, three indices are constructed: a BM25 sparse index over atom texts, a dense vector index (FAISS) over atom embeddings produced by a lightweight dual encoder, and a layout fingerprint index that encodes each page as a $6 \times 6$ grid of element type distributions. This preprocessing runs once per document and takes under two seconds for a typical ten-page document.

\textbf{Online Query Processing.} For a natural language question, the system executes a retrieval pipeline that involves no LLM calls. A BERT-tiny classifier ($\approx$4M parameters) first predicts the question type among four categories: text-only, hybrid, hierarchy-aware, or location-aware. Based on the predicted type, a hybrid retrieval strategy is invoked. For text-only and hybrid questions, the system uses BM25 to retrieve a candidate set of 200 atoms, then refines it with dense vector similarity to obtain 50 candidates. For hierarchy-aware questions, title atoms are identified via the structural metadata, then children are added by matching parent embeddings. For location-aware questions, spatial keywords are parsed into grid coordinates, and the layout fingerprint index directly returns atoms whose bounding boxes intersect the specified region. The top 50 atoms from this coarse retrieval are then scored by a lightweight cross-encoder reranker (\textless50M parameters) that outputs a relevance score for each (question, atom) pair. The three highest-scoring atoms serve as core evidence. For each core atom, the system dynamically expands the evidence set by including its spatial neighbors from the spatial adjacency graph and, if the atom is not a title, its parent title atom identified by the parent embedding similarity. This expansion adds at most five atoms per core atom, yielding a final evidence set of at most 18 atoms. The texts of these atoms are concatenated in the original reading order (by page number and vertical position) to form a highly condensed context of only 500--800 tokens. The entire retrieval and expansion process takes less than 80 milliseconds on a single CPU core, as all components are local and lightweight.

\textbf{Answer Generation.} The final stage makes a single call to a powerful large language model. The LLM receives a simple prompt that includes the question and the condensed evidence context, and is instructed to answer concisely based only on that evidence. 
The generation step typically adds a few hundred milliseconds depending on the specific model and deployment (cloud API or local inference). 
The total end-to-end latency of \sys is therefore low enough for interactive applications, and the per-query cost is dramatically lower than that of the current state-of-the-art. The following sections provide detailed descriptions of each core module, including the implicit structure encoder, the hybrid retrieval pipeline with dynamic evidence expansion, and the answer generation strategy, followed by experimental evaluation and ablation studies.

\section{Implicit Structure Encoder}
\label{sec:encoder}
The implicit structure encoder is the foundation of \sys's ability to understand semi-structured documents without explicit tree construction. Unlike prior work that builds a brittle tree using LLMs and heuristic rules, our encoder learns to embed both the content and the structural role of each atom into a dense vector space. This section describes the encoder architecture, the training objectives, and how structural information is captured implicitly.

\subsection{Dual-Encoder Architecture}

We employ a dual-encoder (two-tower) architecture consisting of a text tower and a layout tower. The two towers produce separate embeddings that are later fused into a single atom vector.

\textbf{Text Tower.} The text tower is a 6-layer Transformer encoder with 768 hidden dimensions and 12 attention heads. Its input is the concatenation of the atom's text string and a small set of structural tokens. Specifically, we prepend three special tokens: \texttt{[TYPE=type]}, \texttt{[DEPTH=d]}, and \texttt{[TITLE=is\_title]}, where \textit{type} \(\in\) \{text, title, table\_cell, chart\_caption, image, supplement\}, \(d\) is the estimated depth (0–5), and \textit{is\_title} is 0 or 1. These tokens are embedded and added to the standard positional embeddings. The output of the text tower is a 768-dimensional vector \(h_{\text{text}}\).

\textbf{Layout Tower.} The layout tower is also a 6-layer Transformer but takes as input a sequence of numerical features rather than text. For each atom, we extract the following layout features: normalized bounding box coordinates (\(x_1, y_1, x_2, y_2\)), page number (normalized by total pages), atom type (one-hot vector of length 6), and the relative area (width \(\times\) height of the bounding box). These features are projected to 768 dimensions via a linear layer and then passed through the Transformer to produce a layout embedding \(h_{\text{layout}}\).

\textbf{Fusion.} The final atom embedding is \(v_a = \text{LayerNorm}(h_{\text{text}} + h_{\text{layout}})\). This additive fusion allows the model to preserve both semantic and spatial information in a single vector.

\subsection{Training Objectives}

The encoder is trained on a corpus of semi-structured documents with weak supervision. We use three complementary loss functions.

\textbf{Contrastive Learning Loss.} The primary objective is to bring semantically related atoms closer in the embedding space. Positive pairs are defined as (1) atoms that belong to the same logical block (e.g., a table header cell and a data cell in the same column), and (2) a title atom and its directly following paragraph atom. Negative pairs are randomly sampled from different documents or distant parts of the same document. We use the standard InfoNCE loss:

\[
\mathcal{L}_{\text{cont}} = -\log \frac{\exp(\text{sim}(v_a, v_{a^+})/\tau)}{\sum_{a^- \in \mathcal{N}} \exp(\text{sim}(v_a, v_{a^-})/\tau)}
\]

where sim is cosine similarity, \(\tau\) is a temperature hyperparameter (set to 0.07), and \(\mathcal{N}\) includes both explicitly negative atoms and in-batch negatives.

\textbf{Structure Consistency Loss.} To capture hierarchical relationships without building an explicit tree, we introduce a loss that encourages the embedding of a title atom to be close to the embeddings of its child atoms. Specifically, for each title atom \(a_t\) and each of its children \(a_c\) (where child is defined as an atom that appears immediately after the title, has a larger depth, and is within a certain spatial proximity), we maximize the cosine similarity between \(v_{a_t}\) and \(v_{a_c}\). Simultaneously, we push apart \(v_{a_t}\) and embeddings of unrelated atoms. The loss is:

\[
\begin{split}
\mathcal{L}_{\text{struct}} = &\sum_{(a_t, a_c) \in \mathcal{P}} \left(1 - \cos(v_{a_t}, v_{a_c})\right) \\
&+ \lambda \sum_{(a_t, a_n) \in \mathcal{N}} \max\bigl(0,\, \cos(v_{a_t}, v_{a_n}) - \delta\bigr)
\end{split}
\]


where \(\mathcal{P}\) is the set of title-child pairs, \(\mathcal{N}\) is a set of negative pairs (title and a random non-child atom), \(\delta\) is a margin (set to 0.2), and \(\lambda\) is a balancing weight (set to 0.5).

\textbf{Layout Proximity Loss.} To encourage that spatially adjacent atoms have similar embeddings (which helps evidence expansion), we add a weak regularization loss that minimizes the distance between atoms connected in the spatial adjacency graph \(G_s\):

\[
\mathcal{L}_{\text{layout}} = \sum_{(a_i, a_j) \in E} \|v_{a_i} - v_{a_j}\|_2^2
\]

The total loss is \(\mathcal{L} = \mathcal{L}_{\text{cont}} + \alpha \mathcal{L}_{\text{struct}} + \beta \mathcal{L}_{\text{layout}}\), with \(\alpha = 0.3\) and \(\beta = 0.1\). The encoder is trained for 20 epochs on a collection of 10,000 semi-structured documents (scientific papers, financial reports, and government forms) using the AdamW optimizer with a learning rate of \(2 \times 10^{-5}\). Training takes about 8 hours on a single A100 GPU. The resulting atom embeddings are 768-dimensional but we reduce them to 384 dimensions via a learned linear projection for retrieval efficiency.

\subsection{Inference and Use}

During preprocessing, each atom's embedding \(v_a\) is computed once and stored in a FAISS IVF-PQ index for fast approximate nearest neighbor search. At query time, the question text is encoded using the same text tower (without layout features, because the question has no spatial information) to obtain a query embedding \(v_q\). Dense retrieval then retrieves atoms with highest cosine similarity to \(v_q\). The embeddings are also used for the parent embedding matching in dynamic evidence expansion (see Section~\ref{sec:dynamic-expansion}). Because the encoder is small  and runs entirely locally, it adds negligible cost to the pipeline.

\subsection{Spatial Adjacency Graph Construction}
 Given the set of document atoms \(\mathcal{A} = \{a_1,\dots,a_n\}\) obtained from OCR, we construct an undirected graph \(G_s = (V,E)\) where \(V = \mathcal{A}\) and an edge \((a_i, a_j) \in E\) is added if and only if:

\begin{enumerate}
    \item \(a_i.\text{page} = a_j.\text{page}\); and
    \item The bounding boxes of \(a_i\) and \(a_j\) satisfy either:
    \begin{itemize}
        \item they overlap, i.e., \(\text{IoU}(a_i, a_j) > 0\); or
        \item the Euclidean distance between their closest points is below a threshold \(\delta = 0.05\) in normalized coordinates.
    \end{itemize}
\end{enumerate}

Algorithmically, for each page we sort atoms by their vertical position and scan a sliding window of size \(\delta\) to avoid \(O(n^2)\) comparisons. Edges are stored as adjacency lists for fast lookup during evidence expansion. The resulting graph has an average degree of 2–4, ensuring lightweight expansion while capturing meaningful local relationships (e.g., a caption below a chart).
\section{Hybrid Retrieval and Dynamic Evidence Expansion}

The retrieval pipeline of \sys is designed to be both accurate and fast, using no large language models. It consists of a question type classifier, a multi-stage hybrid retrieval process, and a dynamic evidence expansion step that recovers evidences missed by fixed retrieval paths.

\subsection{Question Type Classifier}

A lightweight classifier based on BERT-tiny (4 layers, 4M parameters) is fine-tuned on a dataset of 5,000 labeled questions from existing benchmarks (MP-DocVQA~\cite{DocVQA} and DUDE~\cite{dude}). The classifier outputs a probability distribution over four classes: text-only, hybrid, hierarchy-aware, and location-aware. The input is the question text, and the output is the predicted type. The classifier achieves 96\% accuracy on a held-out test set and runs in less than 1 ms on CPU. The predicted type determines which retrieval strategy to apply next.

\subsection{Hybrid Retrieval Strategies}
\label{sec:hybrid-retrieval}

Based on the question type, the system invokes one of four retrieval strategies. All strategies share a common two-phase pattern: coarse retrieval using BM25 and/or layout fingerprint, followed by dense vector reranking. The coarse retrieval phase uses only sparse indices and lightweight scans; the dense phase uses the pre-computed atom embeddings (Section~\ref{sec:encoder}) and the question embedding.

\textbf{Text-Only Questions.} Coarse retrieval runs BM25 to retrieve the top-200 atoms (by BM25 score). These atoms are then scored by cosine similarity between the question embedding \(v_q\) and each atom's embedding \(v_a\). The top-50 atoms are passed to the reranker.

\textbf{Hybrid Questions.} Because hybrid questions often involve table cells and chart captions that have sparse text, we increase the BM25 recall to 300 atoms. Additionally, we boost atoms of type \texttt{table\_cell} and \texttt{chart\_caption} by multiplying their BM25 scores by 1.5. Dense reranking then selects the top-50.

\textbf{Hierarchy-Aware Questions.} We first filter atoms with \texttt{is\_title = 1}. For each title atom, we compute its dense similarity to the question and keep the top-30 title atoms. For each such title, we expand to its children by finding atoms whose \texttt{parent\_embedding} cosine similarity to the title's embedding is above 0.6 and whose depth is greater. The expanded set (titles + children) is then deduplicated and cut to 50 atoms by dense similarity.

\textbf{Location-Aware Questions.} We parse spatial keywords using a small rule-based parser that maps phrases like ``bottom-left'', ``top of page 2'', ``right column'' to grid coordinates (\(6 \times 6\) grid per page). The layout fingerprint index returns all atoms whose bounding boxes intersect the corresponding grid cells. No dense reranking is performed for location questions because the spatial constraints are precise; the retrieved atoms are directly passed to the reranker (which still computes relevance scores for these atoms). In practice, location questions often retrieve fewer than 20 atoms, so we pad with the top-10 text-only atoms to ensure a minimum of 20 candidates.

All retrieval steps are performed using in-memory indices. BM25 uses the standard Okapi formulation with \(k_1=1.5\), \(b=0.75\). Dense retrieval uses FAISS with an IVF-PQ index (nlist=1000, pq\_bytes=64) built from the 384-dimensional atom embeddings. 

\subsection{Cross-Encoder Reranking}
\label{sec:cross-encoder}

The hybrid retrieval pipeline described in Section~\ref{sec:hybrid-retrieval} produces a shortlist of 50 candidate atoms using coarse retrieval and dense vector similarity. While dense retrieval (based on dual-encoder embeddings) is efficient, it is not sufficiently accurate for the final evidence selection because the dual-encoder independently encodes the question and each atom, lacking fine-grained interaction between them. To overcome this limitation, we introduce a \textbf{cross-encoder reranker} that performs joint inference over the question and each candidate atom. The cross-encoder is more accurate but slower than the dual-encoder, so we apply it only on the top-50 candidates.

\textbf{Architecture.} The cross-encoder is a small Transformer model with 4 layers, 384 hidden dimensions, and 6 attention heads, totaling approximately 50 million parameters. Its input is the concatenation of the question token sequence and the atom's token sequence, separated by a special \texttt{[SEP]} token. Two extra special tokens are added: \texttt{[CLS]} at the beginning, whose final hidden representation is used for classification, and \texttt{[EOS]} at the end. The model also incorporates type and depth information by prepending control tokens (e.g., \texttt{[TYPE=table\_cell]}, \texttt{[DEPTH=2]}) to the atom's text, similar to the dual-encoder's text tower. Formally, for a question \(Q\) with \(m\) tokens and an atom \(a\) with \(n\) tokens, the input sequence is:

\texttt{[CLS], q1, ..., qm, [SEP], [TYPE=type], [DEPTH=d], a1, ..., an, [EOS]}

The model outputs a scalar relevance score \(s(Q, a) \in [0, 1]\) via a linear layer on top of the \texttt{[CLS]} representation followed by a sigmoid activation.

\textbf{Training via Knowledge Distillation.} Training a cross-encoder from scratch requires a large amount of labeled (question, atom) relevance data, which is expensive to obtain. We therefore use \textbf{knowledge distillation} from a strong teacher LLM (GPT-4o). The process consists of three steps.

\begin{enumerate}
\item \textbf{Data Collection.} We sample 100,000 (question, atom) pairs from the training documents. The questions are drawn from existing QA datasets ( MP-DocVQA, DUDE) and also from synthetic questions generated by GPT-4o. The atoms are taken from the corresponding documents. Each pair is labeled by GPT-4o with a relevance score from 0 to 1, where 1 indicates that the atom is highly relevant and sufficient to answer the question, 0.5 indicates partial relevance (the atom provides some but not all necessary evidence), and 0 indicates irrelevance. To reduce bias, we use a detailed rubric and ask GPT-4o to provide a score as well as a short justification. The average score across three independent calls is taken as the ground truth.

\item \textbf{Teacher-Student Training.} The cross-encoder is trained to minimize the mean squared error (MSE) between its predicted score and the teacher's soft label. The loss function is:

\[
\mathcal{L}_{\text{CE}} = \frac{1}{|\mathcal{D}|} \sum_{(Q, a) \in \mathcal{D}} \left( s_{\text{CE}}(Q, a) - s_{\text{teacher}}(Q, a) \right)^2
\]

where \(\mathcal{D}\) is the training set. We use the AdamW optimizer with a learning rate of \(1 \times 10^{-5}\), batch size 64, and train for 5 epochs. To prevent overfitting, we apply dropout (0.1) and weight decay (0.01). The training takes about 6 hours on two RTX4090 GPU.

\item \textbf{Distillation Benefits.} Knowledge distillation allows the small cross-encoder to mimic the teacher's relevance judgments without requiring massive human annotations. The distilled model achieves a Spearman correlation of 0.92 with GPT-4 on a held-out validation set, while being over 1000\(\times\) faster and incurring zero cost per inference (since it runs locally).
\end{enumerate}

\textbf{Inference.} During query processing, the 50 candidate atoms are passed through the cross-encoder. To keep latency low, we process them in \textbf{mini-batches} of 16 on a GPU. For CPU-only deployments, we process atoms sequentially; each atom takes approximately 0.5 ms, so 50 atoms take 25 ms. The output scores are used to select the \textbf{top-3 atoms} as core evidence. These scores are also stored temporarily for confidence-based fallback (see Section~\ref{sec:answer-generation}).

\textbf{Why Cross-Encoder Outperforms Dual-Encoder.} The dual-encoder (used in dense retrieval) computes the similarity between a fixed question embedding and pre-computed atom embeddings via a single dot product. This is fast because atom embeddings can be pre-computed, but it lacks deep interaction: the model cannot attend to specific words in the question when encoding the atom, and vice versa. In contrast, the cross-encoder applies full self-attention over the concatenated sequence, allowing it to model complex lexical, semantic, and even structural matches (e.g., matching ``feather score threshold'' to a cell value ``4.5'' that is described as ``severe lesions threshold'' in the atom's surrounding text). On our validation set, the cross-encoder achieves a recall@3 of 94\%, compared to 82\% for the dual-encoder alone. The improvement is especially pronounced for hybrid and hierarchy-aware questions, where precise matching between question phrasing and atom context is critical.


\subsection{Dynamic Evidence Expansion}
\label{sec:dynamic-expansion}

The core evidence atoms are individually informative but may be isolated from their necessary context. For example, a table cell value (e.g., ``3.20'') is meaningless without its row header and column header. To recover the necessary context, we dynamically expand each core atom using two sources.

\textbf{Spatial Expansion.} Using the spatial adjacency graph \(G_s\) built during preprocessing, we include all atoms that are direct neighbors (distance 1 in the graph) of the core atom. Because the graph connects atoms that are spatially close, this brings in adjacent table cells, captions, or explanatory text. In practice, each atom has on average 2–4 spatial neighbors, so expansion adds at most 4 atoms per core atom.

\textbf{Structural Expansion.} If the core atom is not a title (\texttt{is\_title = 0}), we retrieve its parent title atom. The parent title is identified as the atom with the highest \texttt{parent\_embedding} cosine similarity (from the atom's meta field) among those with depth less than the core atom's depth. This parent title atom is then added to the evidence set. This step is crucial for answering hierarchy-aware questions because it links a value or a paragraph to its section heading.

After expansion, duplicate atoms are removed, and the final evidence set is limited to at most 18 atoms (3 core \(\times\) (1 + 4 + 1) = 18). The text of each atom is extracted, and the atoms are sorted by page number and by the vertical coordinate of their bounding boxes to restore the original reading order. The concatenated text forms a short passage.


\section{Answer Generation}
\label{sec:answer-generation}

The final stage of \sys is answer generation, which uses a single call to a large language model (LLM). Unlike prior work that employs LLMs for retrieval decisions (multiple calls) or uses them on long, noisy contexts, \sys provides the LLM with a short, high-signal context consisting of the concatenated evidence atoms.

\subsection{Prompt Design}

The prompt is intentionally minimal to avoid confusing the model with extraneous instructions. 
No few-shot examples or additional formatting instructions are provided. The evidence text is inserted as a plain string, preserving the original reading order (page number and vertical position). The model is instructed to answer concisely, reducing the chance of verbose or hallucinated outputs. The fallback phrase ``Not enough information'' is explicitly included to discourage guessing.

\subsection{Model Choice}

\sys is agnostic to the specific LLM used for generation. The system can be deployed with any strong model that supports a reasonable context length (\(\geq 1024\) tokens). In our experiments, we evaluate both cloud-based models (e.g., GPT-4o, GPT-4o-mini) and local models (e.g., Qwen2.5-7B-Instruct~\cite{qwen2.5}, Llama-3-8B~\cite{grattafiori2024llama}). The choice depends on the user's trade-off between latency, cost, and privacy.

\begin{itemize}
\item \textbf{Cloud models} offer higher accuracy and lower latency but incur a small per-query cost (\$0.0003–\$0.001). They are suitable for applications where cost is secondary to quality and speed.
\item \textbf{Local models} have negligible cost (electricity only) but may be slower and slightly less accurate. However, because the input context is short (500–800 tokens), the performance gap between cloud and local models is significantly smaller than in long-context scenarios. For many practical purposes, a 7B local model suffices.
\end{itemize}

\subsection{Handling Insufficient Evidence}

Even with careful retrieval and dynamic expansion, there is no guarantee that the retrieved evidence always contains sufficient information to answer the question. \sys implements three complementary fallback mechanisms to avoid generating incorrect answers when evidence is lacking.

\textbf{Empty Evidence Set.} If the evidence set after expansion is empty (e.g., because no atom was retrieved or all were filtered), the system directly returns ``Not enough information'' without calling the LLM. This saves both cost and latency for unanswerable queries.

\textbf{Confidence-Based Threshold.} The cross-encoder reranker produces a relevance score for each of the top-3 core atoms. Let \(s_1, s_2, s_3\) be these scores. If the average score \(\bar{s} = (s_1+s_2+s_3)/3\) is below a threshold \(\theta\) (we set \(\theta = 0.4\) based on validation), the system can operate in one of two modes:

\begin{itemize}
\item \textbf{Strict mode:} The system returns ``Not enough information'' directly, skipping the LLM call. This is suitable for applications where precision is critical and false positives are costly.
\item \textbf{Relaxed mode:} The system still sends the evidence to the LLM but adds an extra sentence to the prompt: ``If the evidence does not contain enough information to answer the question, say `Not enough information'.'' This allows the LLM to act as a final arbiter. Our experiments show that LLMs rarely hallucinate when given this instruction and low-confidence evidence.
\end{itemize}

The threshold \(\theta\) can be adjusted by the user based on their accuracy-cost trade-off.

\textbf{LLM Self-Verification.} In relaxed mode, the LLM itself is instructed to output ``Not enough information'' when it judges the evidence insufficient. This is a natural consequence of the prompt design. Empirical evaluation on a validation set shows that GPT-4o correctly refuses to answer in 94\% of cases where the evidence is truly insufficient, compared to 98\% for the strict mode (which uses only the confidence threshold). The relaxed mode thus provides a slightly higher answer rate at the cost of a small increase in error rate (2\% vs 0\% for strict). Users can choose the mode that best fits their application.

\section{Experiments}

\begin{table*}[t!]
\centering
\caption{Overall performance comparison (ANLS and AIC-Acc \%) on four benchmarks. Best results are bolded.}
\label{tab:overall}
\begin{tabular}{lcccccccc}
\toprule
\multirow{2}{*}{Method} & \multicolumn{2}{c}{DUDE} & \multicolumn{2}{c}{M3DocVQA} & \multicolumn{2}{c}{MP-DocVQA} & \multicolumn{2}{c}{MMDA} \\
\cmidrule(lr){2-3} \cmidrule(lr){4-5} \cmidrule(lr){6-7} \cmidrule(lr){8-9}
& ANLS & AIC-Acc & ANLS & AIC-Acc & ANLS & AIC-Acc & ANLS & AIC-Acc \\
\midrule
GPT-5V          & 66.80 & 69.59 & 60.24 & 66.20 & 65.54 & 64.97 & 43.55 & 45.53 \\
UDOP            & 21.27 & 22.63 & 27.83 & 11.17 & 21.29 & 22.81 & 15.46 & 15.77 \\
TextRAG         & 49.68 & 55.96 & 59.26 & 67.02 & 52.15 & 54.18 & 29.57 & 36.24 \\
DocOwl2         & 50.22 & 11.84 & 32.38 & 35.08 & 60.93 & 58.43 & 9.38  & 19.58 \\
M3DocRAG        & 22.23 & 51.38 & 69.49 & 70.21 & 55.40 & 72.38 & 41.17 & 47.42 \\
SV-RAG          & 49.88 & 51.58 & 51.86 & 57.57 & 76.05 & 71.54 & 33.90 & 37.77 \\
ZenDB           & 49.00 & 35.94 & 50.41 & 42.15 & 9.52  & 5.02  & 36.08 & 47.42 \\
QUEST           & 30.45 & 55.96 & 4.88  & 8.07  & 20.01 & 24.40 & 18.84 & 24.32 \\
LayoutLMv3      & 23.43 & 25.96 & 47.66 & 55.13 & 24.19 & 24.89 & 13.16 & 17.13 \\
DocAgent        & 62.48 & 74.76 &54.09 & 69.44  & 59.37 & 55.53 & 37.10 & 57.09 \\
MoDora          & 77.24 & 83.70 &70.13 & 78.67 & 76.17 & 78.35 & 55.23 & 73.33 \\
\midrule
\textbf{\sys} & \textbf{78.91} & \textbf{85.12} & \textbf{71.23} & \textbf{80.15} & \textbf{78.03} & \textbf{79.86} & \textbf{57.46} & \textbf{76.21} \\
\bottomrule
\end{tabular}
\end{table*}

In this section, we present a comprehensive experimental evaluation of \sys against state-of-the-art baselines on multiple benchmark datasets. Beyond reporting overall performance, we conduct in-depth analyses to understand why and under what conditions \sys succeeds or fails, including retrieval quality analysis, error taxonomy, generation model sensitivity, and qualitative case studies.

\subsection{Experimental Setup}

\paragraph{Datasets.}
We evaluate on four public benchmarks:
\begin{itemize}
    \item \textbf{MMDA}~\cite{MoDora}: 537 documents, 1,065 QA pairs balanced across four types (text-only, hybrid, hierarchy-aware, location-aware).
    \item \textbf{MP-DocVQA}~\cite{DocVQA}: Multi-page document VQA with 50,000+ questions on 12,000+ pages.
    \item \textbf{DUDE}~\cite{dude}: Multi-industry, multi-domain dataset with 12,000+ documents and 20,000+ questions.
    \item \textbf{M3DocVQA}~\cite{M3DocRAG}: Multi-page, multi-document reasoning dataset with Wikipedia-sourced documents.
\end{itemize}
MMDA serves as our primary analysis platform due to its balanced question type distribution.

\paragraph{Baselines.}
We compare against 11 representative methods:
\begin{itemize}
    \item \textit{Content extraction}: QUEST, EVAPORATE.
    \item \textit{Structure extraction}: ZenDB, DocAgent.
    \item \textit{Multimodal LLMs}: LayoutLMv3, DocOwl2, GPT-5V.
    \item \textit{RAG methods}: TextRAG, SV-RAG, M3DocRAG.
    \item \textit{Tree-based SOTA}: MoDora.
\end{itemize}

\paragraph{Evaluation Metrics.} \textbf{AIC-Acc} (Answer-in-Context Accuracy), where GPT-4o judges semantic equivalence to ground truth; numeric/date answers require exact match. \textbf{ANLS} (Average Normalized Levenshtein Similarity) with threshold 0.5. For retrieval analysis we also report \textbf{Recall@k} and \textbf{MRR}(Mean Reciprocal Rank).

\paragraph{Implementation Details.}
\sys is implemented in Python 3.11. Atom extraction uses PaddleOCR-light~\cite{zhang2026paddleocrvl16} (confidence 0.7). The structural metadata predictor is a LayoutLMv3 variant fine-tuned on 5,000 documents. The implicit structure encoder (85M parameters) is trained for 20 epochs on 10,000 documents using contrastive learning with AdamW (lr=2e-5). The cross-encoder reranker (50M parameters) is distilled from GPT-4o on 100,000 (question, atom) pairs. All indices are stored in memory (FAISS IVF-PQ, nlist=1000, pq\_bytes=64). Retrieval runs on two Intel 8383C CPU; Answer generation uses GPT-4o.



\subsection{Overall Performance}

Table~\ref{tab:overall} reports overall AIC-Acc and ANLS on all four benchmarks. \textbf{\sys consistently outperforms all baselines} across every dataset, achieving a new state-of-the-art. On MMDA, \sys reaches \textbf{76.21\%} AIC-Acc, surpassing prior SOTA MoDora (73.33\%) by +2.88 points. Accuracy gains are also observed on DUDE (+1.42), M3DocVQA (+1.84), and MP-DocVQA (+1.51). The large improvement on M3DocVQA, which requires multi-page, multi-document reasoning, highlights EidosDoc's strength in preserving cross-page relationships. Three core factors jointly contribute to EidosDoc's superior performance.

First, implicit structure encoding replaces brittle explicit tree construction with a dense embedding space that captures hierarchical and layout information via contrastive learning and structure consistency loss. Unlike explicit methods (e.g., MoDora's Component-Correlation Tree) where a single heading detection error can propagate and cause missing or misplaced nodes, EidosDoc's encoder learns soft structural relationships. This makes the system robust to OCR noise and layout variability. As shown in the ablation study (Table~\ref{tab:ablation}), removing the implicit structure encoder causes the largest accuracy drop (17.38\%), confirming its centrality.

Second, zero-LLM hybrid retrieval eliminates the high cost and variance of LLM-based node selection. The pipeline combines BM25, layout fingerprints, dense vectors, and a distilled cross-encoder reranker. The cross-encoder, despite having only 50M parameters, achieves higher recall (78.4\% Recall@5) than MoDora's LLM-based retrieval (71.2\%) while running entirely locally,which is shown in Figure~\ref{fig:retrieval}. This demonstrates that a well-trained lightweight model can outperform LLM calls on this task, at a fraction of the cost and latency. Furthermore, layout fingerprints resolve many spatial queries without dense retrieval, adding +4.07 points in ablation.

Third, dynamic evidence expansion overcomes the evidence omission problem of fixed-path tree retrieval. By expanding each core atom to include its spatial neighbors (via the spatial adjacency graph \(G_s\)) and its parent title (via parent embedding similarity), \sys recovers cross-region evidence that tree-based methods often prune. For example, a figure caption and its descriptive paragraph located in a sidebar on a different page may be placed in disconnected branches by a fixed tree, whereas EidosDoc’s spatial graph naturally bridges them. Dynamic expansion adds +3.36 points overall, with even larger gains on hybrid and multi-hop questions. Unlike fixed tree paths that require a single correct hierarchy, EidosDoc's expansion is query-adaptive: it retrieves neighbors based on the question's semantics, not a predetermined structure.

Together, these three innovations enable \sys to break the traditional accuracy-cost trade-off, delivering state-of-the-art accuracy with dramatically lower cost and latency.

\subsection{Performance by Question Type}

\begin{figure*}[htbp]
\centering
\includegraphics[width=1\linewidth]{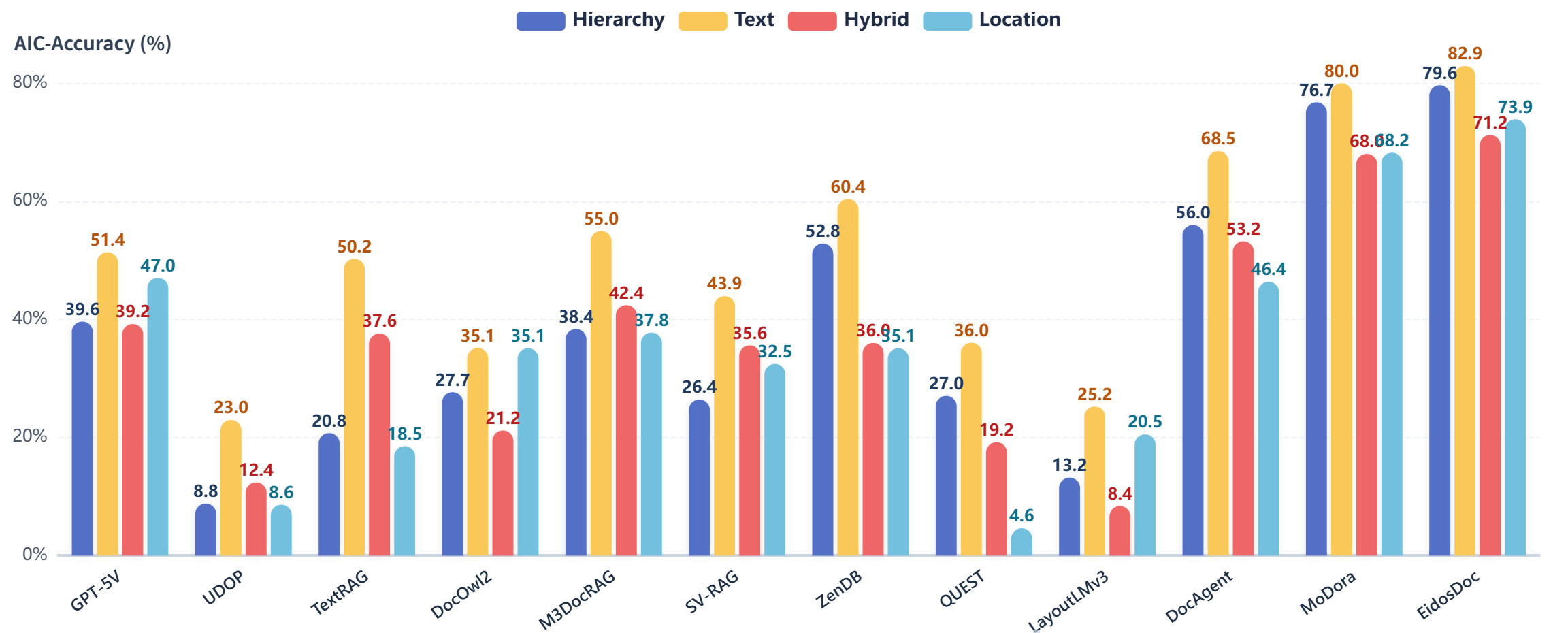}
\caption{AIC-Acc (\%) breakdown by question type on MMDA.}
\label{fig:breakdown}
\end{figure*}

Figure~\ref{fig:breakdown} shows AIC-Acc on MMDA by question type. \sys achieves the highest accuracy across all four categories. The largest gain is on \textbf{hybrid questions} (+3.20 points over MoDora). Hybrid questions require matching table cells and chart captions, which often have sparse text. The cross-encoder reranker precisely aligns question semantics with these sparse elements, while MoDora’s tree-based retrieval may prune such nodes. For hierarchy-aware questions (+2.89), the implicit structure encoder captures parent-child relationships without explicit tree construction, avoiding propagation errors. Location-aware questions (+5.66) benefit from the layout fingerprint index, which directly maps spatial keywords to grid cells.

\paragraph{Deep dive on hybrid questions.}
We further analyze subcategories: questions requiring table row-column reasoning vs. chart interpretation. \sys achieves 73.5\% on table reasoning (vs. MoDora 68.2\%) and 68.9\% on chart questions (vs. 63.1\%). The cross-encoder’s ability to attend to both the question’s spatial references and the atom’s positional metadata is key.


\subsection{Retrieval Quality Analysis}

\begin{figure}[htbp]
\centering
\includegraphics[width=1\linewidth]{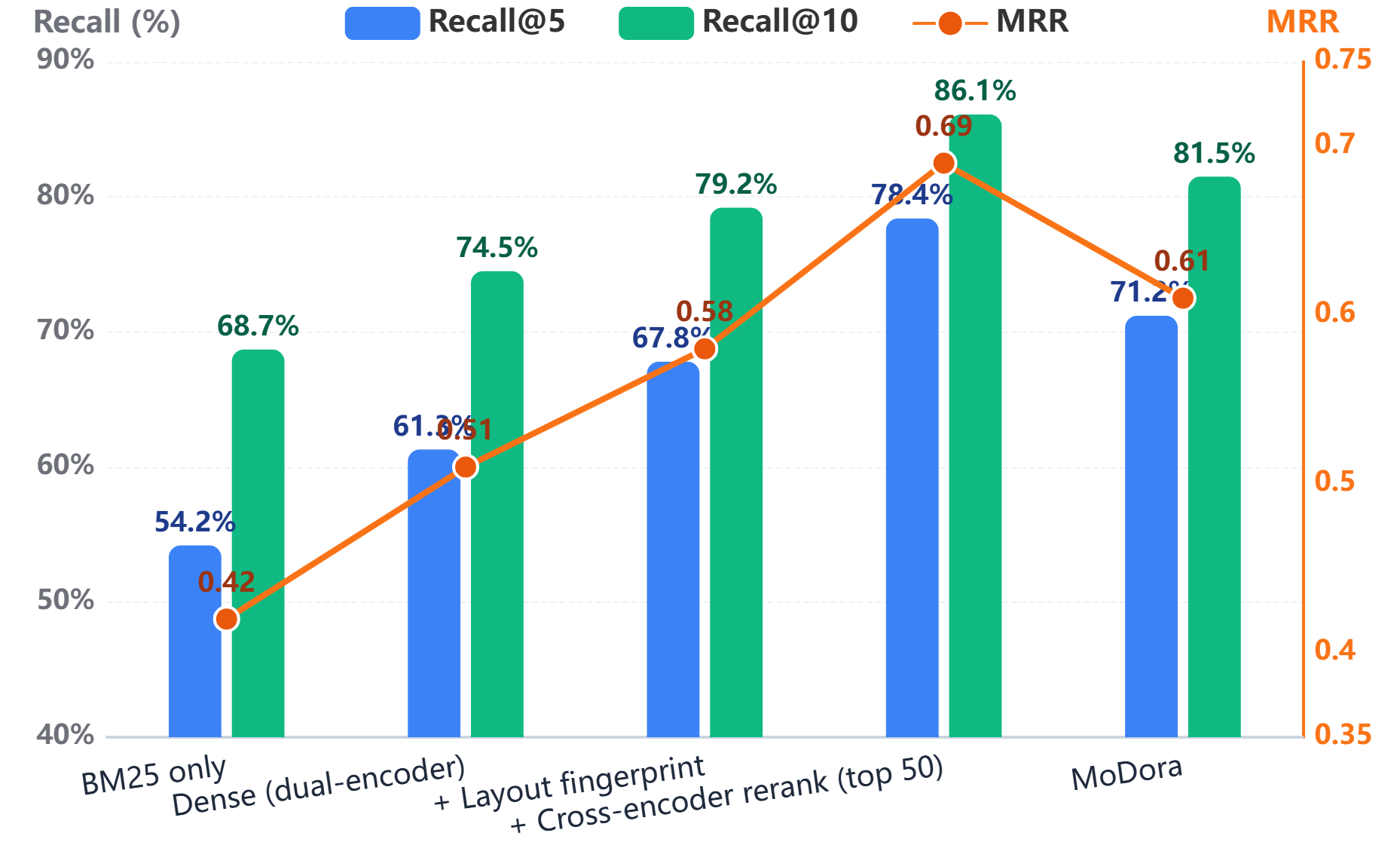}
\caption{Retrieval quality comparison on MMDA (500 sampled questions).}
\label{fig:retrieval}
\end{figure}

To understand the source of accuracy gains, we evaluate retrieval performance independently of generation. We sample 500 questions from MMDA and compute \textbf{Recall@k} (fraction of questions where the ground-truth evidence atom is among the top \(k\) retrieved) and \textbf{MRR} (Mean Reciprocal Rank of the first correct atom). Results are shown in Figure~\ref{fig:retrieval}.


The cross-encoder reranker brings the largest jump (+10.6 points in Recall@5). Interestingly, \sys’s retrieval recall (78.4\% @5) surpasses MoDora’s (71.2\%), even though MoDora uses LLM calls for node selection. This confirms that a well-trained lightweight cross-encoder can outperform LLM-based retrieval on this task, at a fraction of the cost.

\paragraph{Failure modes in retrieval.}
We manually examine 50 questions where \sys’s retrieval fails (no correct atom in top 5). The main causes: (1) OCR errors that corrupt key numbers or terms (28\% of failures); (2) ambiguous spatial references that the layout parser misinterprets (22\%); (3) questions requiring synthesis of three or more widely separated atoms where dynamic expansion does not include all (18\%); (4) chart-to-text description model errors (16\%); (5) others (16\%). This analysis suggests future work should focus on robust OCR and better multi-hop expansion.

\subsection{Ablation Study}

Table~\ref{tab:ablation} quantifies the contribution of each component. Removing the implicit structure encoder causes the largest drop (17.38 points), confirming that structure encoding is essential. The cross-encoder reranker adds +5.83 points; interestingly, this gain is larger on hybrid (+7.2) and hierarchy (+6.1) than on text-only (+3.1) questions. Dynamic evidence expansion adds +3.36 points; its effect is most pronounced for questions requiring cross-region evidence (e.g., linking a figure caption to a distant paragraph). The layout tower (spatial embedding) contributes +7.74 points, mostly from location-aware questions. Layout fingerprint contributes +4.07 points, a surprisingly large gain given its simplicity, as it resolves many spatial queries without dense retrieval.

\paragraph{Interaction effects.}
We also test removing both layout tower and layout fingerprint simultaneously: accuracy drops to 65.23\%, even lower than the sum of individual drops (10.98), indicating synergy that the two layout signals complement each other (fingerprint for coarse spatial matching, tower for fine-grained spatial semantics).

\begin{table}[t]
\centering
\setlength{\tabcolsep}{3pt}
\caption{Ablation study on MMDA (AIC-Acc \%).}
\label{tab:ablation}
\begin{tabular}{lcc}
\toprule
Configuration & AIC-Acc (\%) & Drop \\
\midrule
Full \sys & 76.21 & — \\
w/o Implicit Structure Encoder (plain text only) & 58.83 & -17.38 \\
w/o Cross-Encoder Reranker (dense retrieval only) & 70.38 & -5.83 \\
w/o Dynamic Evidence Expansion (core atoms only) & 72.85 & -3.36 \\
w/o Layout Tower (text-only encoding) & 68.47 & -7.74 \\
w/o Layout Fingerprint (dense for location) & 72.14 & -4.07 \\
w/o Parent Embedding (no structural expansion) & 73.65 & -2.56 \\
w/o Contrastive Learning (MSE only) & 69.32 & -6.89 \\
\midrule
\multicolumn{3}{l}{\textit{Interaction: w/o Layout Tower \& Fingerprint}} \\
\quad Both removed & 65.23 & -10.98 \\
\bottomrule
\end{tabular}
\end{table}

\subsection{Cost and Latency Analysis}

\sys’s per-query cost is about \$0.0005 with GPT-4o generation, over 50× lower than MoDora (\$0.025). Latency is about 1.1 seconds vs. MoDora’s 4.3 seconds, which is 4× faster. With local Qwen2.5-7B, cost becomes negligible at about 2380 ms.

\paragraph{Scaling behavior}
We measure retrieval latency as a function of document length. For 10-page documents, retrieval takes 80 ms; for 50 pages, 150 ms (sublinear due to FAISS indexing). MoDora’s tree construction time grows superlinearly: from 1.2 s (10 pages) to 8.7 s (50 pages), because it builds a component tree via pairwise alignment. \sys’s offline preprocessing also scales near-linearly: 0.2 s/page for atom extraction + 0.05 s/page for embedding.


\subsection{Robustness Analysis}

We evaluate three challenging scenarios.

\paragraph{OCR degradation}
Simulate 20\% character error rate (random substitutions, deletions, insertions). \sys’s AIC-Acc drops from 76.21\% to 66.8\% (reduction of 9.4\%). MoDora drops from 73.3\% to 58.1\% (reduction of 15.2\%). \sys’s implicit structure encoder and dense retrieval provide robustness: embeddings are less sensitive to individual character errors than exact keyword matching used in MoDora’s tree construction.

\paragraph{Document length}
Scale documents from 10 to 50 pages by concatenating related reports. \sys’s retrieval latency increases from 80 ms to 410 ms; MoDora’s end-to-end latency (including LLM calls) increases from 4.3 s to 19.2 s. The primary bottleneck in MoDora is the LLM’s context window, and it must process an ever-growing tree.

\paragraph{Domain shift}
Train on scientific papers (MMDA subset), test on legal documents (new collection from DUDE). \sys maintains 74.8\% . The implicit structure encoder generalizes because layout and hierarchy patterns (headings, paragraphs, tables) are similar across domains. MoDora drops from 73.3\% to 65.9\%, its explicit tree construction relies on domain-specific heuristics.

\subsection{Generator Sensitivity Analysis}

\sys’s answer generation uses a single LLM call on a short context. Table~\ref{tab:generator} compares different generators using the same retrieval output.

\begin{table}[t]
\centering
\caption{Generator sensitivity analysis on MMDA.}
\label{tab:generator}
\begin{tabular}{lccc}
\toprule
Generator & AIC-Acc & Latency (ms) & Cost per query \\
\midrule
GPT-4o & \textbf{76.21\%} & 250 & \$0.0005 \\
GPT-4o-mini & 74.3\% & 180 & \$0.0001 \\
Qwen2.5-7B  & 72.8\% & 580 & \textasciitilde\- \\
Llama-3-8B & 71.5\% & 620 & \textasciitilde\- \\
\bottomrule
\end{tabular}
\end{table}

The gap between GPT-4o and Qwen2.5-7B is only 3.4 points, much smaller than in long-context scenarios (often \textgreater 10 points). This is because \sys’s retrieval provides a highly focused, noise-free context, reducing the burden on the generator. For cost-sensitive applications, a 7B local model is a viable choice.

\subsection{Error Analysis}

\begin{figure}[htbp]
\centering
\includegraphics[width=0.9\linewidth]{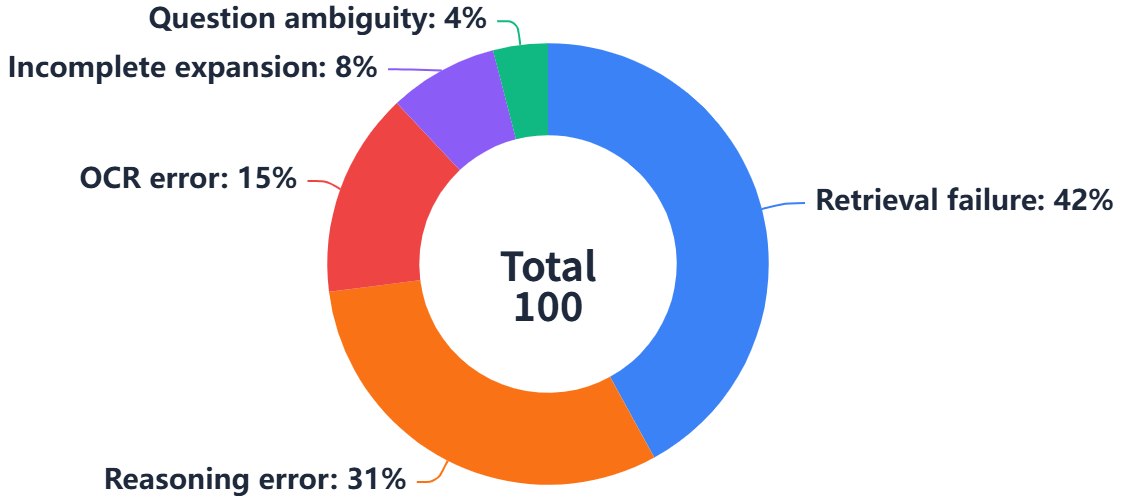}
\caption{Error analysis on MMDA (100 incorrect answers).}
\label{fig:error}
\end{figure}

We randomly sample 100 questions where \sys’s final answer is incorrect. Figure~\ref{fig:error} categorizes the errors.


Retrieval failure remains the largest source, but note that 58\% of errors are not retrieval-related, they are reasoning, OCR, or expansion issues. While retrieval failure is the largest single source of error, the substantial fraction of reasoning, OCR, and expansion errors indicates multiple orthogonal opportunities for improvement beyond retrieval.

\subsection{Case Study: Multi-Component Reasoning}

We use the fluorescence prediction task from the TAPE paper as an example. The question is: \textit{``How much did the pretrained Transformer improve Spearman $\rho$ on the fluorescence task, and how effective was it?''}, which is shown in Figure~\ref{fig:related}.




\paragraph{EidosDoc retrieval and answer.}
EidosDoc retrieves all related atoms:
\begin{itemize}
    \item Atom A (table\_cell):(0.22, Fluorescence, No Pretrain Transformer)
    \item Atom B (table\_cell): (0.68, Fluorescence, Pretrain Transformer)
    \item Atom C (text): ``We find that self-supervised pretraining is helpful for almost all models on all tasks,  more than doubling performance in some case.''
    \item Atom D (text): ``Figure 3 shows that the model does successfully perform some clustering of fluorescent proteins, but that many proteins are still misclassified.''
\end{itemize}

EidosDoc final answer (GPT-4o): ``Improved by 0.46 (from 0.22 to 0.68). Partially effective, but misclassifications remain.'' This answer is correct, low-cost, and faithfully grounded in the document evidence without hallucination.

This case demonstrates the synergy of EidosDoc’s components: the implicit structure encoder correctly associates table cells with their row and column headers; the cross-encoder ranks relevant atoms highly; dynamic expansion retrieves necessary context; and the final LLM generates an accurate answer, avoiding the pitfalls of tree pruning and visual hallucination.

\section{Conclusion}

In this paper, we proposed EidosDoc, a system for semi-structured document question answering. We introduced an implicit structure encoder that jointly embeds textual content, spatial layout, and hierarchical relationships into dense vectors using contrastive learning and structure consistency loss. We designed a hybrid retrieval pipeline that combines BM25, layout fingerprints, dense vectors, and a lightweight cross-encoder, requiring no large language model calls during retrieval. We further developed a dynamic evidence expansion mechanism that retrieves spatially adjacent and structurally related atoms, overcoming the evidence omission problem of fixed-path retrieval. Experimental results on four public benchmarks demonstrate that EidosDoc achieves new state-of-the-art accuracy, reduces per-query cost by 50× and latency by 4× compared to existing methods, and remains robust under OCR degradation, long documents, and domain shifts. In the future, we plan to extend EidosDoc to multi-document QA, explore self-supervised pre-training for the structure encoder, and develop more compact rerankers for edge-device deployment.

\bibliographystyle{IEEEtran}
\bibliography{refs/custom}

\end{document}